\documentclass[12pt]{article}
\usepackage[T2A]{fontenc}
\usepackage[utf8]{inputenc}
\usepackage[english]{babel}
\usepackage{amssymb}
\usepackage{amsmath}
\usepackage{graphicx}
\usepackage[small,sc,center]{titlesec}
\usepackage{indentfirst}
\usepackage{siunitx}

\newcommand{\nc}{\newcommand}
\nc{\lhe}{L_\mathrm{He}}
\nc{\lhemax}{L_\mathrm{He,max}}
\nc{\mhe}{M_\mathrm{He}}
\nc{\mzams}{M_\mathrm{ZAMS}}
\nc{\rph}{R_\mathrm{ph}}
\nc{\tev}{t_\mathrm{ev}}
\begin{document}

\begin{center}
\textbf{Radial pulsations of red giant branch stars}

\vskip 3mm
\textbf{\quad Yu. A. Fadeyev\footnote{E--mail: fadeyev@inasan.ru}}

\textit{Institute of Astronomy, Russian Academy of Sciences, Pyatnitskaya ul. 48, Moscow, 119017 Russia} \\

Received May 15, 2017
\end{center}

\textbf{Abstract} ---
We performed hydrodynamic computations of nonlinear stellar pulsations of
population I stars at the evolutionary stages of the ascending red giant branch
and the following luminosity drop due to the core helium flash.
Red giants populating this region of the Hertzsprung--Russel diagram were found
to be the fundamental mode pulsators.
The pulsation period is the largest at the tip of the red giant branch and for
stars with initial masses from $1.1M_\odot$ to $1.9M_\odot$ ranges from
$\Pi\approx 254$ day to $\Pi\approx 33$ day, respectively.
The rate of period change during the core helium flash is comparable with rates
of secular period change in Mira type variables during the thermal pulse
in the helium shell source.
The period change rate is largest ($\dot\Pi/\Pi\approx -10^{-2}\ \text{yr}^{-1}$)
in stars with initial mass $\mzams=1.1M_\odot$ and decreases to
$\dot\Pi/\Pi\sim -10^{-3}\ \text{yr}^{-1}$
for stars of the evolutionary sequence $\mzams=1.9M_\odot$.
Theoretical light curves of red giants pulsating with periods $\Pi > 200$
show the presence of the secondary maximum similar to that observed in many Miras.

Keywords: \textit{stars: variable and peculiar}

\newpage
\section{introduction}

The red giant branch (RGB) of the Hertzsprung--Russel diagram (HRD) is populated
by the post--main--sequence stars with the hydrogen shell source surrounding the
inert helium core.
The central gas density of the red giant with mass $M\le 2M_\odot$ is as high as
$\rho_\mathrm{c}\sim 10^6~\text{g}\ \text{cm}^{-3}$ whereas the central temperature
is $T_\mathrm{c}\lesssim 8\times 10^7$~K, so that the electron gas of the helium core
is in degenerate state.
Due to the weak temperature dependence of the pressure of a degenerate gas
the gradual increase of the central temperature in the contracting degenerate
helium core leads to the helium flash, that is the explosive growth of energy
generation by the triple--alpha reactions
(Schwarzschild and H\"arm 1962; H\"arm and Schwarzschild 1964; Thomas 1967).
The energy released during the helium flash is highest in low--mass stars
and decreases with increasing mass of the star, that is with decreasing degeneracy
of the electron gas.
For example, in population II red giants with mass $M=0.6M_\odot$ the maximum
luminosity generated in the helium core is
$\lhemax\approx 7\times 10^{10}L_\odot$ (Despain 1981),
whereas for $M=2M_\odot$ becomes as low as $\lhemax\approx 1.6\times 10^7 L_\odot$
(Kippenhahn et al. 2012).
The short--term increase of the helium core luminosity is responsible for cease
of energy production in the hydrogen shell source and decrease of the surface
luminosity  by one or two orders of magnitude.
Therefore, the stars at the evolutionary stage preceding the helium flash populate
the tip of the red giant branch (TRGB).

Interest in TRGB stars was confined mainly to low--mass ($M\approx 0.6M_\odot$)
population II red giants due to the fact that their luminosity is almost
independent of the star age and is only a function of the mass of the degenerate
helium core (Salaris et al. 2002; Greggio and Renzini 2011).
Therefore, measurements of the brightness of TRGB stars allow us to determine
the distances for galaxies of the Local Group with accuracy comparable with that
based on the Cepheid period--luminosity relation (Lee et al. 1993).
However one should bear in mind that a disadvantage of the distance scale based
on the photometry of TRGB stars is due to uncertainties in bolometric
corrections of red giants (Salaris and Cassisi 1997).
These difficulties might, however, be avoided in the case of observed pulsational
variability because the stellar radius can be determined by methods of stellar
pulsation theory.

By this time, the pulsational variability is detected in RGB stars of the local
solar neighbourhood (Tabur et al. 2009) and Magellanic Clouds (Ita et al. 2002).
The goal of the present study is to investigate the evolutionary and
pulsation models of population I RGB stars with initial mass fractional abundances
of helium and heavier elements $Y=0.28$ and $Z=0.02$, respectively.
The initial masses of evolutionary models are $1.1M_\odot\le\mzams\le 1.9M_\odot$.
The results presented below are based on consistent stellar evolution and
stellar pulsation computations.
The methods of computations are described in our previous papers (Fadeyev 2016; 2017).

\section{results of computations}

\begin{figure}
\centerline{\includegraphics[width=12cm]{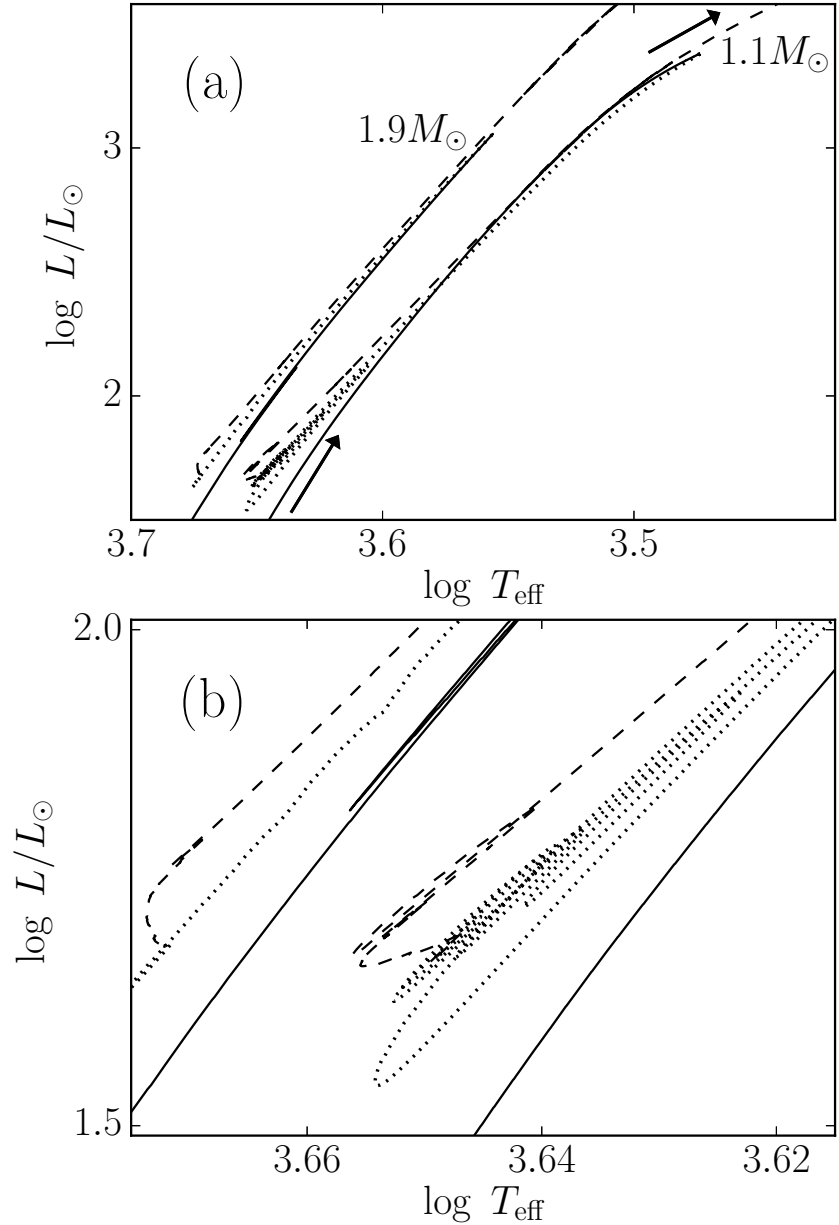}}
\caption{(а) -- Evolutionary tracks of RGB stars with initial masses $1.1M_\odot$
         and $1.9M_\odot$ on the HRD.
         The values of $\mzams$ are indicated near the TRGB points.
         The solid, dashed and dotted lines indicate the evolutionary stages
         corresponding to the ascending branch of red giants, the helium flash
         and AGB, respectively.
         Arrows in the lower and upper parts of the figure indicate the
         direction of evolution during the RGB and early AGB stages, respectively.
         (б) -- Evolutionary tracks near the minimum of the surface luminosity
         after the helium flash.
         }
\label{fig1}
\end{figure}

Evolutionary tracks of the stars with initial masses $\mzams=1.1M_\odot$ and
$\mzams=1.9M_\odot$ in the RGB region are shown in Fig.~\ref{fig1}(a).
Each track is represented by the solid line (the ascending RGB), the dotted line
(the stage of helium flash) and the dashed line (the early AGB phase).
The initial masses are indicated near TRGB points and the direction of evolution
is indicated by arrows.
In the present work we investigated radial oscillations of red giants during
ascent of the RGB and then during the helium flash.
On the HRD the evolutionary tracks of these stages almost coincide, albeit
the evolutionary time differ by several orders of magnitude.
For the sake of more clarity, the zoomed evolutionary tracks near the minimum
surface luminosity are shown in Fig.~\ref{fig1}(b).
Red giants are the fundamental mode pulsators and their oiscillations are driven
by the $\kappa$--mechanism in the hydrogen ionization zone.

\subsection{models of pulsating stars ascending the red giant branch}

\begin{figure}
\centerline{\includegraphics[width=15cm]{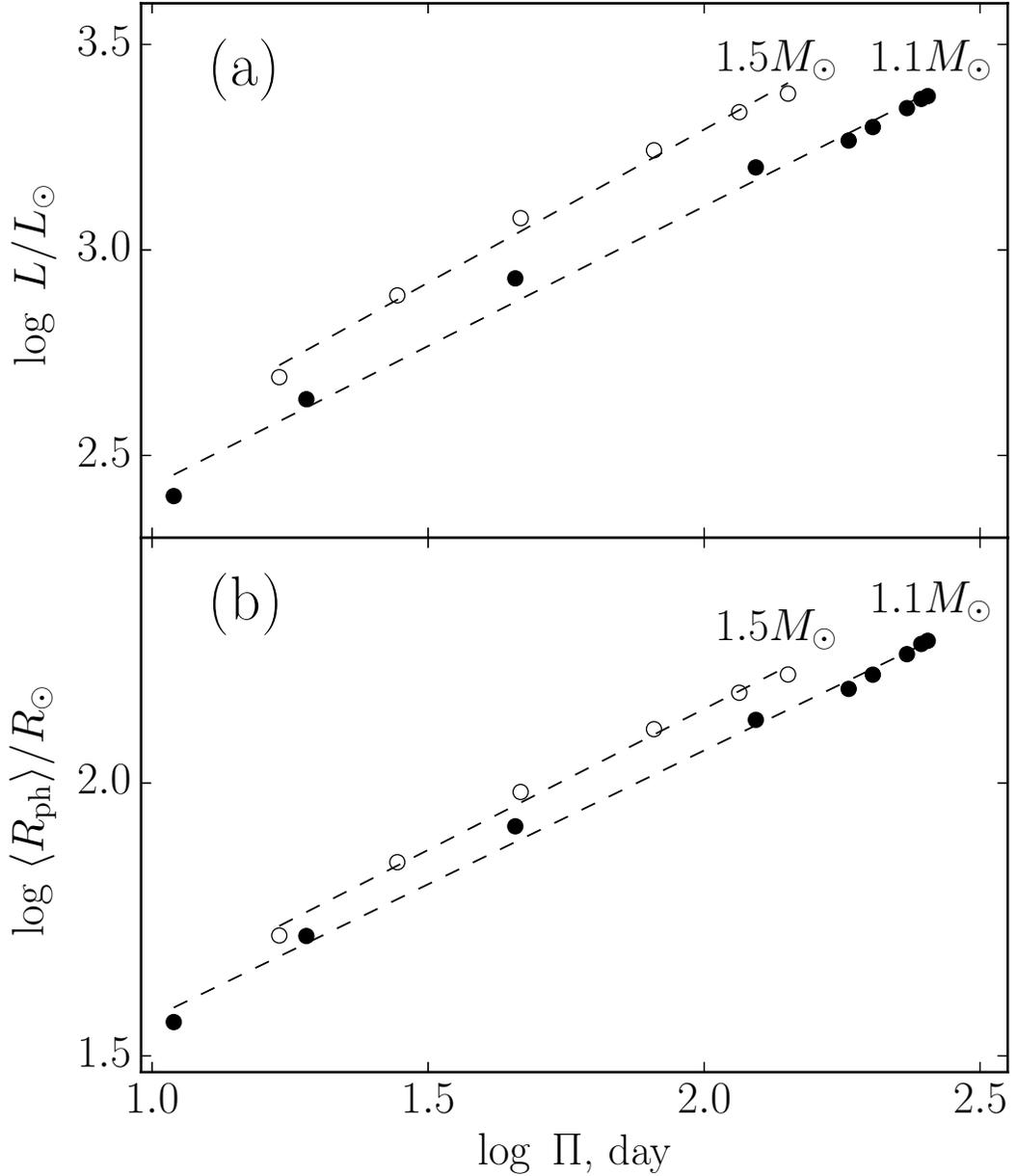}}
\caption{The period--luminosity (a) and period--radius (b) relations of RGB stars
         with initial masses $\mzams=1.1M_\odot$ and $\mzams=1.5M_\odot$.
         The filled and open circles represent the hydrodynamic models.
         The dashed lines show relations ($\ref{p-l}$) and ($\ref{p-r}$).}
\label{fig2}
\end{figure}

Evolution of the star during the hydrogen shell burning phase is accompanied by
the growth of the mass of the degenerate helium core and the surface luminosity
(Refsdal and Weigert 1970).
Therefore  during evolution along the ascending red giant branch the period
of radial oscillations increases more than an order of magnitude, the increase of the
period being more significant in stars with smaller masses due to the lower gas density
of the outer convection zone.
Evolutionary changes of pulsational properties of RGB stars with initial masses
$\mzams=1.1M_\odot$ and $\mzams=1.5M_\odot$ are illustrated by the period--luminosity
and period--radius relations shown in Fig.~\ref{fig2}.
One should be noted that for the period--radius relation we used the mean values
of the photosphere radius $\langle\rph\rangle$ obtained from the limit cycle
hydrodynamic models.

The least--squares fits for the period--luminosity and period--radius relations
are written as
\begin{equation}
\label{p-l}
 \log L/L_\odot = \left\{
\begin{array}{ll}
 1.747 + 0.680 \log\Pi, \qquad &  (\mzams=1.1M_\odot) ,
   \\
 1.803 + 0.745 \log\Pi,        &  (\mzams=1.5M_\odot) .
\end{array}
\right.
\end{equation}
и
\begin{equation}
\label{p-r}
 \log \langle\rph\rangle/R_\odot = \left\{
\begin{array}{ll}
 1.078 + 0.491 \log\Pi, \qquad &  (\mzams=1.1M_\odot) ,
   \\
 1.102 + 0.517 \log\Pi,        &  (\mzams=1.5M_\odot) ,
\end{array}
\right.
\end{equation}
respectively.

TRGB stars are of greatest interest and their evolutionary and pulsational
properties are listed in the table,
where $M$, $t$ and $L$ are the mass, the age and the luminosity of the model of the
evolutionary sequence, $\lhemax$ is the maximum luminosity of the helium core
during the core helium flash, $\mhe$ is the mass of the helium core,
$\Pi$ and $Q$ are the pulsation period and the pulsation constant.
For stars with initial masses $1.1M_\odot\le \mzams\le 1.9M_\odot$
the pulsation constant can be fitted as a function of the initial mass
$\mzams$ with relative deviation less than 2\% by the following relation:
\begin{equation}
Q = 0.1130 - 0.0300 \mzams/M_\odot .
\end{equation}

\begin{figure}
\centerline{\includegraphics[width=15cm]{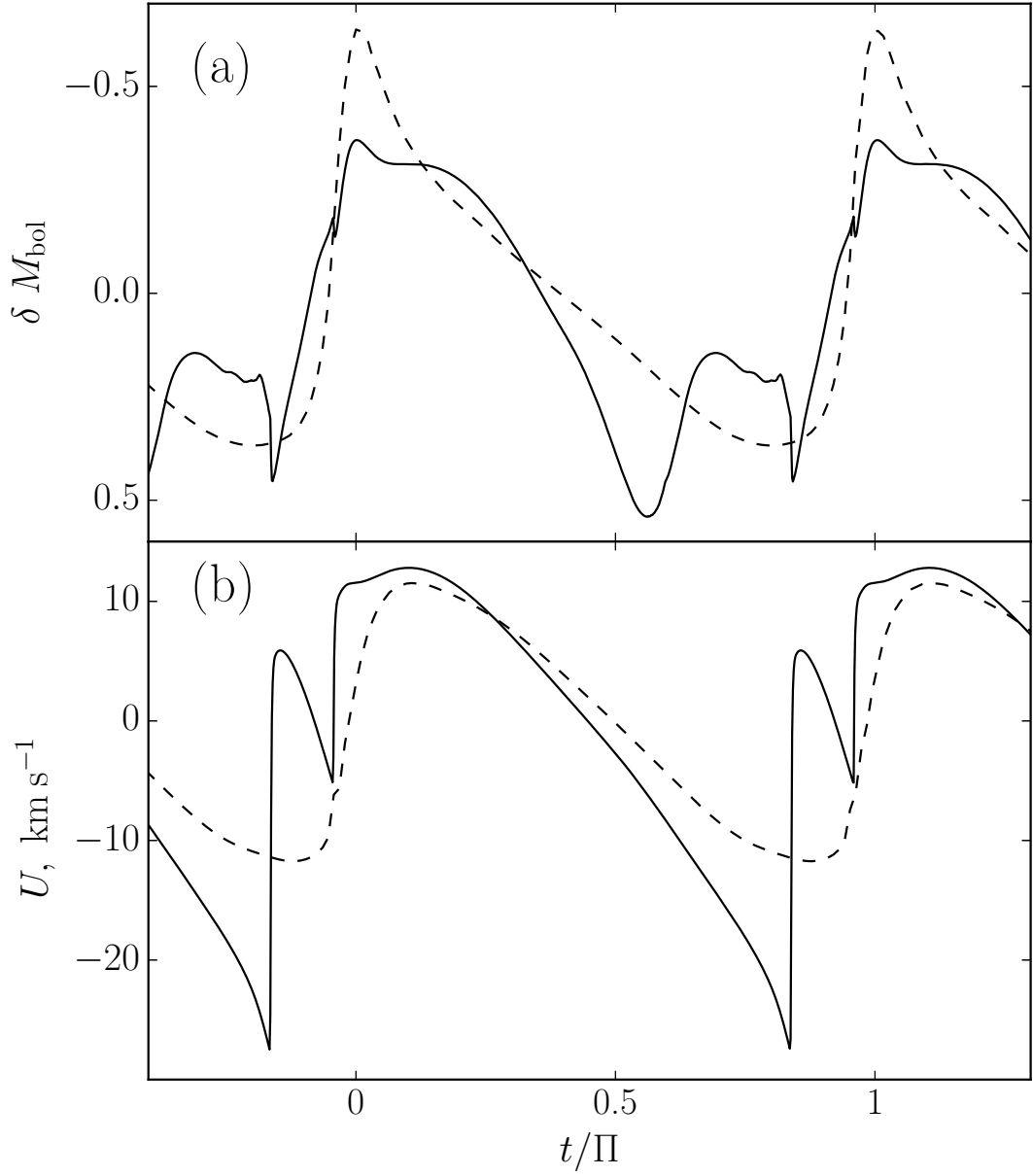}}
\caption{Temporal dependencies of the bolometric light (а) and the gas flow velocity
         on the outer boundary (b) of the hydrodynamic models of TRGB stars with
         initial masses $1.1M_\odot$ (solid lines) and $1.9M_\odot$ (dashed lines)
         with pulsation periods 254 and 33.4 day, respectively.}
\label{fig3}
\end{figure}

As seen from table, the fundamental mode period decreases by a factor of eight
as the initial mass increases from $1.1M_\odot$ to $1.9M_\odot$.
Moreover, the light and the outer boundary velocity curves approach a sinusoidal
shape with increasing stellar mass, whereas  the amplitude of
limit cycle oscillations remains almost unchanged (see Fig.~\ref{fig3}).
The light and velocity curves of the low--mass model ($\mzams=1.1M_\odot$)
deserve attention due to the existence of the secondary maximum (the hump)
which is usually considered as the typical feature of Mira type variables
(Lockwood and Wing 1971) which are the stars at a more advanced AGB evolutionary
stage.

\subsection{stellar pulsations during the core helium flash}

\begin{figure}
\centerline{\includegraphics[width=15cm]{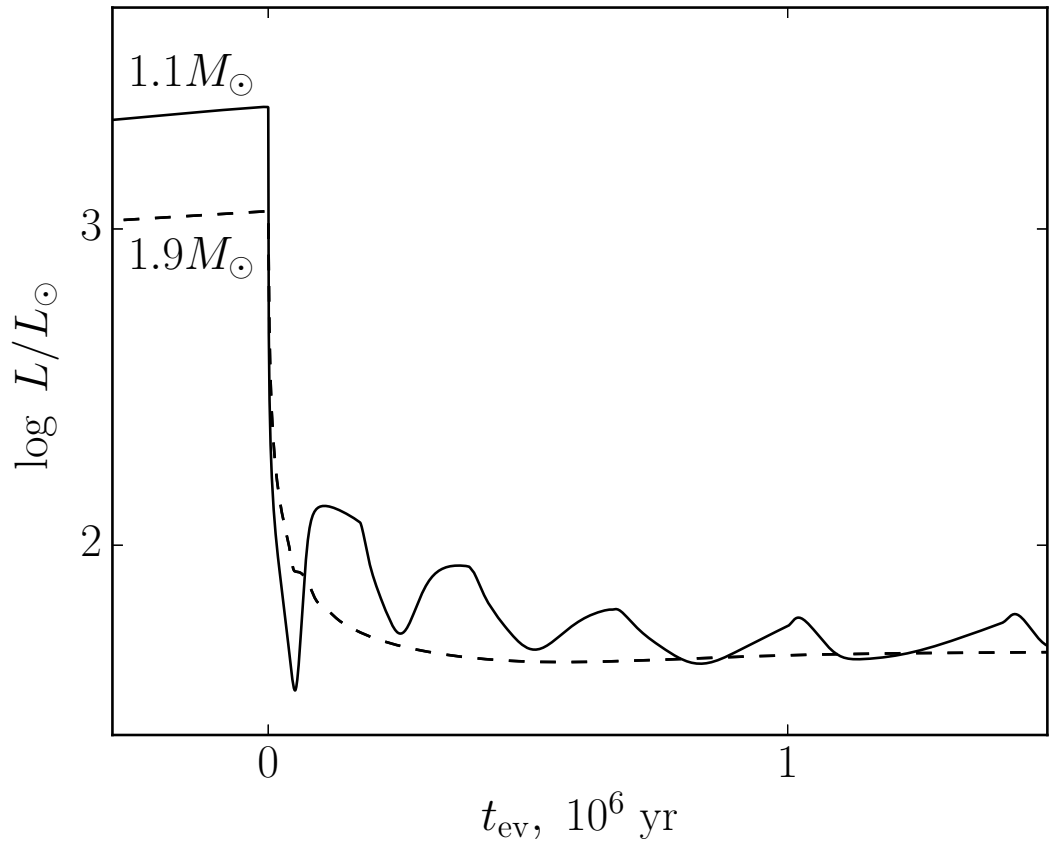}}
\caption{The time variation of the surface luminosity $L$ after the core helium flash
         ($\tev>0$) in stars with initial masses $\mzams=1.1M_\odot$ and
         $\mzams=1.9M_\odot$.}
\label{fig4}
\end{figure}

Fig.~\ref{fig4} shows the temporal dependencies of the surface luminosity $L$
after the core helium flash in red giants with initial masses $\mzams=1.1M_\odot$
and $\mzams=1.9M_\odot$.
For the sake of convenience the evolutionary time $\tev$ is set to zero at the
maximum luminosity of the helium core.
The initial conditions of hydrodynamic models were determined for $\tev > 0$.

One should be noted that the core helium flash occurs as a sequence of peaks
of the helium core luminosity with gradually decreasing amplitude,
i.e. relaxation oscillations (Schwarzschild and H\"arm 1967; Thomas 1967).
Relaxation oscillations are most conspicuous in low--mass red giants and are
responsible for oscillations of the surface luminosity (see Fig.~\ref{fig4}).
The present study was aimed to determine the rate of period change during the
gradual drop of the surface luminosity, so that the stage of relaxation oscillations
after the helium flash ($\tev > 10^4$~yr for the evolutionary sequence $\mzams=1.1 M_\odot$)
was ignored.

\begin{figure}
\centerline{\includegraphics[width=15cm]{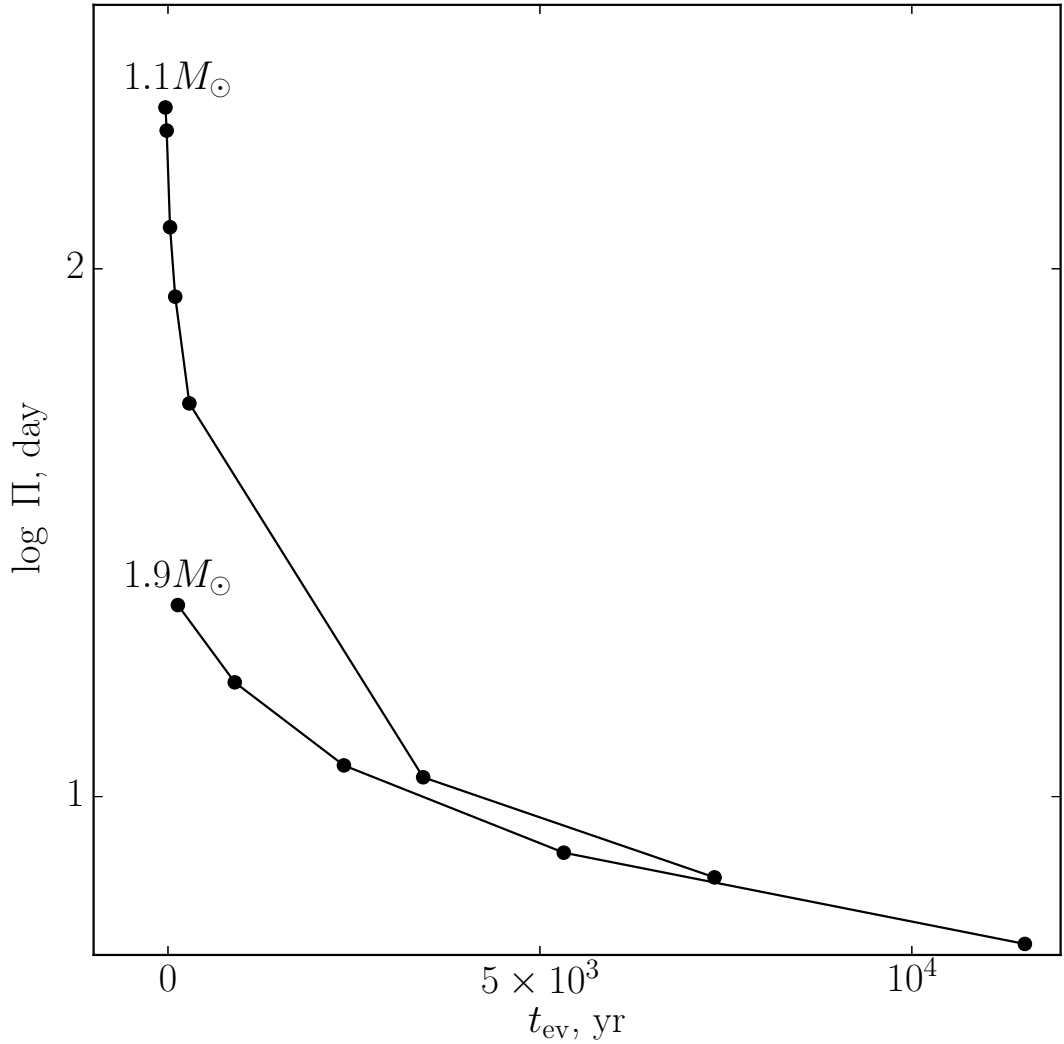}}
\caption{The period of radial pulsations after the maximum luminosity of the
         helium core in stars with initial masses $1.1M_\odot$ and $1.9M_\odot$
         as a function of evolutionary time $\tev$.
         Filled circles represent pulsation periods of hydrodynamic models.}
\label{fig5}
\end{figure}

\begin{figure}
\centerline{\includegraphics[width=15cm]{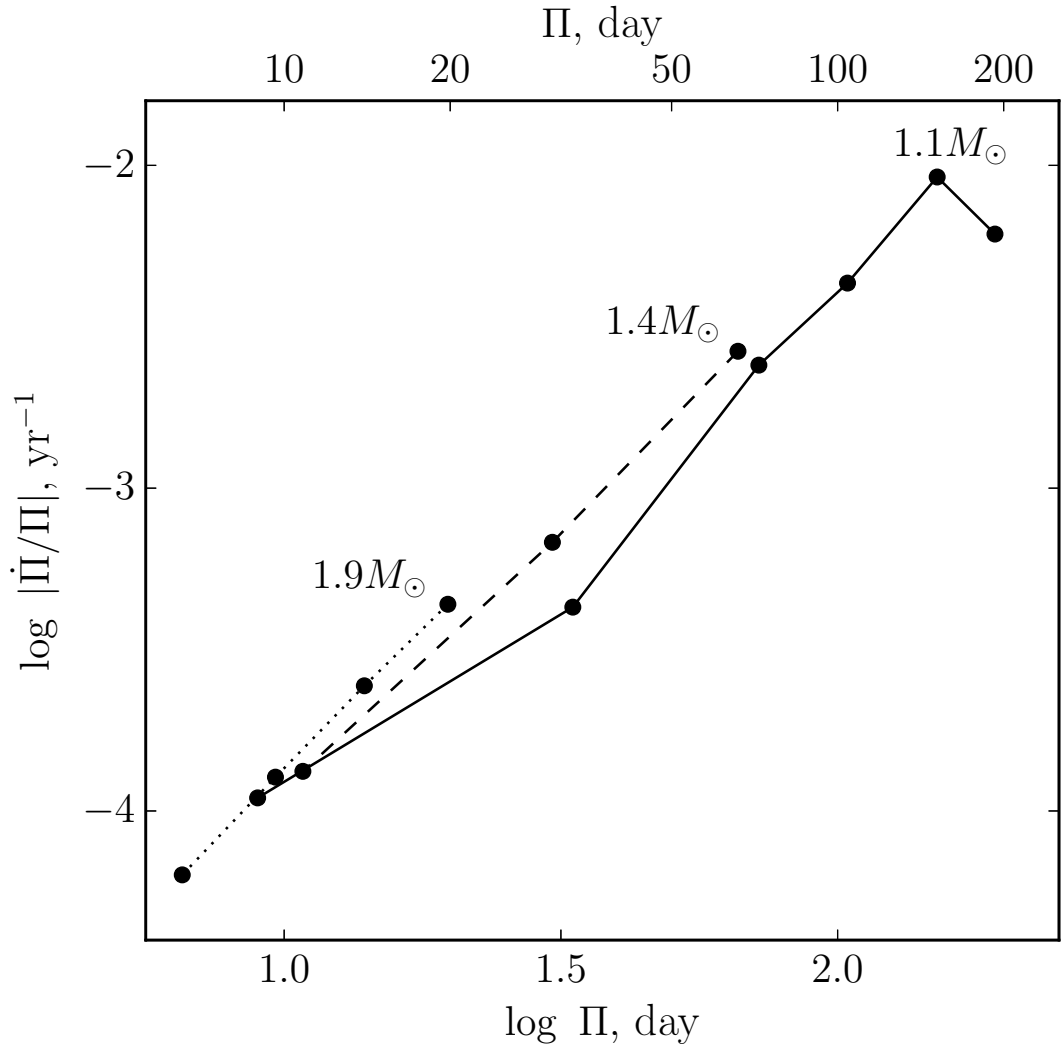}}
\caption{Theoretical estimates of the rate of radial oscillation period change
         as a function of the fundamental mode period
         after the core helium flash in the models of evolutionary sequences with
         initial masses $1.1M_\odot$, $1.4M_\odot$ and $1.9M_\odot$.}
\label{fig6}
\end{figure}

The results of hydrodynamic computations for the models of red giants during the
evolutionary stage of the core helium flash are presented in Fig.~\ref{fig5},
where the plots of the radial pulsation period of stars with initial masses
$1.1M_\odot$ and $1.9M_\odot$ are shown as a function of evolutionary time $\tev$.
Theoretical estimates of the period change rate $\dot\Pi$ were evaluated
by the second--order numerical differentiation.
The diagram period--period change rate for models of evolutionary
sequences with initial masses $\mzams=1.1M_\odot$, $1.4M_\odot$ and $1.9M_\odot$
is shown in Fig.~\ref{fig6}.

\section{discussion}

Results of evolutionary and hydrodynamic computations for models of RGB stars
presented in this study supplement ones from our previous works devoted
to investigation of pulsations in stars of the asymptotic giant branch
(Fadeyev 2016; 2017).
The main conclusion of these works is that the RGB stars as well as
the more luminous AGB stars are the fundamental mode pulsators.

Of great interest is the role of the pulsation period as a criterion
in establishing the evolutionary status of the red giant.
As seen in table, the luminosity of TRGB stars with $\mzams\le 1.4M_\odot$
is almost independent of the initial mass and the star age, whereas  the
period of radial oscillations gradually decreases with increasing $\mzams$.
If we assume that the age of red giants of the local solar neighbourhood
observed by Tabur et al. (2009) is $t \approx 4.7\times 10^9$~yr,
then from table we find that the maximum period of the fundamental mode
is $\Pi\approx 160$ day.
A cursory inspection of Table~3 from the work Tabur et al. (2009) allows
us to conclude that among 261 observed red giants 45 stars have periods
longer than 200 day.
In order to avoid such a contradiction one should either to verify
the methods of period determination or to assume that these red giants
belong AGB stars.
It should be noted that division of red giants into RGB and AGB stars
is complicated due to the fact hydrodynamic models of pulsating RGB stars
show the presence of secondary maxima on their light curves.
Furthermore, the rate of period decrease during the core helium flash is
by an order of magnitude comparable with rates of secular period decrease
during the thermal pulse of typical AGB stars (Fadeyev 2017).

\newpage
\section*{references}

\begin{enumerate}

\item K.H. Despain, Astrophys. J. \textbf{251}, 639 (1981).

\item Yu.A. Fadeyev, Pis'ma Astron. Zh. \textbf{42}, 731 (2016)
      [Astron. Lett. \textbf{42}, 665 (2016)].

\item Yu.A. Fadeyev, Pis'ma Astron. Zh. \textbf{43}, in press (2017)
       [Astron.Lett. \textbf{43}, in press (2017)].

\item L. Greggio and A. Renzini,
      \textit{Stellar populations. A user guide from low to high redshift} (Wiley-VCH, 2011).

\item R. H\"arm and M. Schwarzschild, Astrophys. J. \textbf{139}, 594 (1964).

\item Y. Ita, T. Tanab\'e, N. Matsunaga, Y. Nakajima, C. Nagashima, T. Nagayama, D. Kato,
      M. Kurita, T. Nagata, S. Sato, M. Tamura, H. Nakaya, Y. Nakada, MNRAS \textbf{337}, L31 (2002).

\item R. Kippenhahn, A. Weigert, and A. Weiss,
      \textit{Stellar structure and evolution, 2nd ed.} (Springer, 2012). 

\item M.G. Lee, W.L. Freedman, and B.F. Madore, Astrophys. J. \textbf{417}, 553 (1993).

\item G.W. Lockwood and R.F. Wing, Astrophys. J. \textbf{169}, 63 (1971).

\item S. Refsdal and A. Weigert, 1970, Astron. Astrophys. \textbf{6}, 426 (1970).

\item M. Salaris and S. Cassisi, MNRAS \textbf{289}, 406 (1997).

\item M. Salaris, S. Cassisi, and A. Weiss, Publ. Astron. Soc. Pacific \textbf{114}, 375 (2002).

\item V. Tabur, T.R. Bedding, L.L. Kiss, T.T. Moon, B. Szeidl, and H. Kjeldsen,
      MNRAS \textbf{400}, 1945 (2009).

\item H.--C. Thomas, Zeitschrift f\"ur Astrophys. \textbf{67}, 420 (1967).

\item M. Schwarzschild and R. H\"arm, Astrophys. J \textbf{136}, 158 (1962).

\item M. Schwarzschild and R. H\"arm, Astrophys. J \textbf{150}, 961 (1967).

\end{enumerate}

\newpage
\begin{table}
\caption{Models of TRGB stars} 
\label{table1}
\begin{center}
 \begin{tabular}{ccccccrrrr}
  \hline
  $\mzams/M_\odot$  & $M/M_\odot$ & $t,\ 10^9\ \text{yr}$ & $\log L/L_\odot$ & $\log\lhemax/L_\odot$ & $\mhe/M_\odot$ &
  $\Pi$, day      & $Q$, day\\
  \hline
1.1 &   0.871 & 8.49 & 3.374 &  9.268 &  0.464  &  254.2 & 0.0791 \\
1.2 &   0.999 & 6.24 & 3.385 &  9.276 &  0.464  &  226.3 & 0.0785 \\
1.3 &   1.122 & 4.71 & 3.385 &  9.277 &  0.464  &  162.4 & 0.0721 \\
1.4 &   1.241 & 3.64 & 3.384 &  9.278 &  0.464  &  157.6 & 0.0710 \\
1.5 &   1.357 & 2.86 & 3.380 &  9.272 &  0.465  &  141.9 & 0.0690 \\
1.6 &   1.475 & 2.29 & 3.362 &  9.236 &  0.460  &  116.1 & 0.0660 \\
1.7 &   1.599 & 1.88 & 3.307 &  9.122 &  0.451  &   83.7 & 0.0624 \\
1.8 &   1.728 & 1.58 & 3.207 &  8.824 &  0.433  &   54.3 & 0.0581 \\
1.9 &   1.854 & 1.34 & 3.054 &  8.008 &  0.410  &   33.4 & 0.0555 \\
\hline          
 \end{tabular}
\end{center}
\end{table}
\clearpage

\newpage
\clearpage

\newpage
\clearpage

\newpage
\clearpage

\newpage
\clearpage

\newpage
\clearpage

\newpage
\clearpage

\end{document}